\documentclass[journal]{IEEEtran}

\usepackage{cite}

\ifCLASSINFOpdf
   \usepackage[pdftex]{graphicx}
   \graphicspath{{../pdf/}{../jpeg/}}
   \DeclareGraphicsExtensions{.pdf,.jpeg,.png}
\else
   \usepackage[dvips]{graphicx}
   \graphicspath{{../eps/}}
   \DeclareGraphicsExtensions{.eps}
\fi

\usepackage{amsmath,array,multirow}
\usepackage{amssymb}
\usepackage{algorithm,algorithmicx,algpseudocode}
\usepackage{fixltx2e,stfloats,url}
\usepackage[justification=centering]{caption}
\ifCLASSOPTIONcompsoc
  \usepackage[caption=false,font=normalsize,labelfont=sf,textfont=sf]{subfig}
\else
  \usepackage[caption=false,font=footnotesize]{subfig}
\fi

\ifCLASSOPTIONcaptionsoff
  \usepackage[nomarkers]{endfloat}
 \let\MYoriglatexcaption\caption
 \renewcommand{\caption}[2][\relax]{\MYoriglatexcaption[#2]{#2}}
\fi

 \let\MYorigsubfloat\subfloat
 \renewcommand{\subfloat}[2][\relax]{\MYorigsubfloat[]{#2}}

\begin{document}

\title{Image-based Insider Threat Detection via Geometric Transformation}

\author{Dongyang Li,
        Lin Yang, Hongguang Zhang, Xiaolei Wang, Linru Ma
        and Junchao Xiao% <-this % stops a space
\thanks{Dongyang Li was with College of Command Information System, Army Engineering University of PLA, Nanjing,
	 211101 China e-mail: donyoung\_lee@sina.com.}% <-this % stops a space
\thanks{Lin Yang, Hongguang Zhang, Xiaolei Wang, Linru Ma and Junchao Xiao are with National Key Laboratory of Science and Technology on Information System Security,Institute of System Engineering, Academy of Military Sciences PLA, Beijing, 100039 China.}}
%\thanks{Manuscript received April 19, 2005; revised August 26, 2015.}

% The paper headers
%\markboth{Journal of \LaTeX\ Class Files,~Vol.~14, No.~8, August~2015}%
%{Shell \MakeLowercase{\textit{et al.}}: Bare Demo of IEEEtran.cls for IEEE Journals}

\maketitle

\begin{abstract}
Insider threat detection has been a challenging task over decades, existing approaches generally employ the traditional generative unsupervised learning methods to produce normal user behavior model and detect significant deviations as anomalies. However, such approaches are insufficient in precision and computational complexity. In this paper, we propose a novel insider threat detection method, Image-based Insider Threat Detector via Geometric Transformation (IGT), which converts the unsupervised anomaly detection into supervised image classification task, and therefore the performance can be boosted via computer vision techniques. To illustrate, our IGT uses a novel image-based feature representation of user behavior by transforming audit logs into grayscale images. By applying multiple geometric transformations on these behavior grayscale images, IGT constructs a self-labelled dataset and then train a behavior classifier to detect anomaly in self-supervised manner. The motivation behind our proposed method is that images converted from normal behavior data may contain unique latent features which keep unchanged after geometric transformation, while malicious ones cannot. Experimental results on CERT dataset show IGT outperforms the classical autoencoder-based unsupervised insider threat detection approaches, and improves the instance and user based Area under the Receiver Operating Characteristic Curve (AUROC) by 4\% and 2\%, respectively.
\end{abstract}

% Note that keywords are not normally used for peerreview papers.

%\IEEEpeerreviewmaketitle

\section{Introduction}

\IEEEPARstart {I}{nsider} threat generally refers to the malicious and unintentional actions on the part of insider, always negatively affecting the confidentiality, integrity or availability of the organization’s information system \cite{costa2016insider}. Due to the fact that insiders are usually knowledgeable about the organization’s security mechanisms and authorized to access the system service, insider threat is one of the challenging threats and hardest to detect. As reported in the Insider Threat Annual Report, 67\% of organizations have experienced one or more insider attacks in the last 12 months, and this digit keeps going to increase with increased economic uncertainty \cite{insider2020report}. In a recent survey, insider attacks are shown to account for 25\% of the cybercrime incidents, and 30\% of respondents indicate that the damage and economic loss caused by insider threats are much severe than external one \cite{insider2018survey}. As such, emerging insider threats such as system sabotage, data breach have been recognized as the critical security challenges faced by various institutions and government agencies. Hence it is urgent to develop effective approaches for detecting malicious insiders accurately.

Insider threat detection is very important, but the sensitivity of attack source and the stealthiness of malicious activities make the identification of insiders very challenging. Firstly, the built-in security defense mechanisms against external attack cannot detect them, and it is necessary to establish additional security procedure to discover insider threat. Secondly, the malicious activities caused by insiders represent only a small portion of their overall activities, and most malicious activities are committed in multiple stages over a long period of times. That is, analysts need to use long-term monitoring across a wide range of audit data sources, thereby increasing the burden of detection. Thirdly, the diversity of insider attack and complex role distribution of organization mean that the “one-fit-all” detection model may not exist, and the detection scheme needs to be dictated by actual requirements. Fourthly, organizations can suffer negative effects (e.g. inefficient work behavior) if an innocent user is classified as suspicious, which violate the original intention of deploying security procedure. Last but not least, researchers cannot acquire and use the real insider dataset easily due to the limitations of privacy and reputation protection, making it difficult for them to effectively evaluate the insider threat detection approaches. In brief, high accuracy requirement, excessive burden in handling big data, and lack of real-world dataset are the main challenges for designing an effective insider threat detection mechanism.

Despite the above challenges, the industrial and academic put forward lots of insider threat detection approaches \cite{liu2019insider,tuor2017deep,sharma2020user,senator2013detecting,jiang2019anomaly,liu2018anomaly,chattopadhyay2018scenario}. Since malicious behavior is widely varying, it is impractical to explicitly characterize insider threat. Instead, most solutions tend to build normal user behavior models by means of historical behavior analysis and anomalies are identified as significant deviations from the normal behavior \cite{homoliak2019insight}. In the modeling process, many classical learning algorithms such as support vector machine (SVM), isolation forest (IF), hidden Markov model (HMM) and Bayesian inference can be used as the benchmark for security analysis, and have achieved the remarkable results in practical applications \cite{liu2018detecting}. Apart from the classification algorithms, the audit data sources also play an important role in determining the detection capability and performance. This is because the malicious threat scenarios are usually not limited to a specific behavior domain, but scattered in multiple domains and composed of multiple activities. If a certain activity of malicious scenario is analysed separately, it may even be normal. For example, the device activity “using a removable drive on the office computer after work” is normal behavior, but it can be judged as malicious when combined with the http activity “uploading files to wikileaks.org”. In other words, the judgment of malicious activity should be combined with the specific context, which puts forward new requirements for multi-source data fusion.

According to the data fusion methods, traditional insider threat detection solutions can be classified into two categories. One category deploys multiple sub-detectors and generates the final decision based on the voting mechanism, where each sub-detector only focuses on a specific type of suspicious activity. The other category combines the statistics extracted from all relevant audit data to form feature vectors, and identifies the suspicious activities using various machine learning classification algorithms. However, due to the fact that whether an activity is malicious or not is closely related to the contextual situation, the performance of sub-detector that only targets specific behavior domain is not satisfying. Therefore, we tend to adopt the second insider threat detection scheme in this paper. But this scheme also suffers the following limitations: i) Feature engineering relies on domain knowledge about how an insider attack is characterized. It is not a trial work to define appropriate feature vectors based on potential threat scenarios, and there is still a lot of room for improvement. ii) The traditional shallow machine learning models are unable to obtain the satisfactory precision due to the complexity and heterogeneity of user behavior data. Thus, one aim of the research is to develop a high-precision insider threat detection method with the deep learning model. Moreover, when it comes to practical application, unsupervised learning method is the first choice for researchers. In this regards, the unsupervised anomaly detection methods can also be roughly categorized into two categories: reconstruction-based anomaly score and reconstruction-based representation learning \cite{golan2018deep}. The former assumes that anomalies and non-anomalies have different latent low-dimensional representations, and it will be difficult to compress and reconstruct the anomalies based on a reconstruction model optimized for non-anomalies. Those samples with large reconstruction errors are regarded as anomalies. The latter uses a two-step approach, which firstly learns a compact representation of the data, and then applying density estimation methods on the lower-dimensional representation. Those samples which lie in low-density region are deemed anomalous. However, in this paper we tend not to use the above two methods based on generative component and instead use a completely different approach to achieve unsupervised detection of insider threat.

Inspired by image classification method GeoTransform \cite{golan2018deep}, we find that the unsupervised classification problem can be converted into supervised classification problem by constructing self-labelled dataset. What’s more, this method can improve the classification accuracy while reducing the problem complexity. Meanwhile, the academic community has begun to adopt transfer learning to the domain of cybersecurity in recent years \cite{gayathri2020image,rezende2017malicious}. On this basis, we propose a novel insider threat detection approach named IGT, which is based on image representation and geometric transformations. In accordance with the principle of comparison with historical baseline, IGT constructs individual behavior model for each user, applies the unsupervised classification method based on geometric transformation to the images converted by user behavioral feature vectors, and finally achieves the precise identification of malicious instances and users. Specifically, we extract the user behavioral representation vectors from all the relevant audit data according to the potential malicious scenarios, and convert them to grayscale images. Then we train a multi-class neural classifier for each user over the self-labelled dataset, which is created from the normal instances and their transformed versions, obtained by applying different geometric transformations. In testing phrase, this classifier is applied on transformed instances of the test sample, and the sample with worse classification results will be judged as malicious. The intuition behind our method is that the images converted from normal behavior data may contain some unique latent features compared to malicious data. Besides, it should be mentioned that all the experiments in this paper are based on the CERT public dataset \cite{glasser2013bridging}.

In summary, this paper makes the following contributions:

Firstly, according to the potential threat scenarios and available audit data, we design a more reasonable feature set, which helps get better performance on representing user behavior. The proposed feature is made up primarily of occurrence time, assigned computer and specific activity, and can be subdivided into three types (i.e. week, day, session) based on the aggregation granularity. The experimental data shows that our feature vector has better behavior representational capacity than other existing feature engineering.

Secondly, we propose a novel insider threat detection approach IGT. By converting behavior feature vectors to grayscale images and constructing self-labelled dataset through geometric transformations, IGT converts unsupervised anomaly detection problem into supervised image classification problem, thereby reducing the complexity. To the best of our knowledge, this is the first work to apply the unsupervised classification method with geometric transformation on insider threat detection.

Thirdly, we evaluate IGT on CERT dataset. Our experiment results show that IGT outperforms the classical autoencoder-based classification method, and improves the instance and user-based AUROC by 4\% and 2\%, respectively.

The rest part of this paper is organized as follows. Section II summarizes the related work on insider threat detection. Section III presents the feature extraction method, and designs an unsupervised insider threat detection mechanism based on image representation and geometric transformation as well as the related algorithm. Section IV details the employed dataset, experimental setting, and evaluation results. Finally, we discuss the weakness and future work in Section V, and make a conclusion in Section VI.

\section{Related Work}
Due to its important role in the field of organization security, the insider threat detection has been widely investigated over many decades. On the one hand, in order to prevent military data from being stolen by insiders, the DARPA consecutively released two insider-related projects, ADAMS and CINDER. As their engineering outcome, the PRODIGAL system takes user activity logs as input, and gets good detection performance by constructing flexible dynamic detection architecture \cite{senator2013detecting}. The technical report released by the CERT Insider Threat Center explored the possible manifestation of insider threats and presented the common guiding mitigation and preventive measures \cite{collins2016common}. On the other hand, there are also many excellent surveys and solutions in the academia. Liu et al. systematically reviewed the present studies on insider threat from the perspective of audit data source \cite{liu2018detecting}. Homoliak et al. proposed a structural taxonomy and novel categorization of insider to systematize knowledge in insider threat research \cite{homoliak2019insight}. Hunker et al. believe that the insider threat detection problem cannot be effectively addressed without collective efforts of psychoanalysis, social relationship investigation and anomaly detection technology \cite{hunker2011insiders}. Although we agree with this opinion, it should be noted that the literature related to nonbehavioral factors (i.e. psychometry, emotion) is outside the scope of this paper. Our focus is how to predict if an employee is behaving abnormally either with respect to his past activity at any given time instance. In this scope, this section will introduce the related studies from the perspective of feature engineering and unsupervised anomaly detection.

\textbf{\emph{1) Feature Extraction.}} Textual audit data such as host logs cannot be directly used in anomaly detection algorithms, so feature extraction is a necessary step to convert them into the numerical vectors. Depending on whether existing human intervention, the feature extraction methods can be classified into two types: statistical features based on artificial definition and hidden features based on representation learning. The former is an intelligible common approach and its core idea is to artificially define indicators that may be related to insider threat as feature attributes by means of expert domain knowledge. These indicators include a variety of types, such as frequency and statistic. Tuor et al. generate 408 features to characterize the behavior pattern by combining different users, time frame and activity frequency information, and it is proved to present good performance \cite{tuor2017deep}. Under the guidance of this combination idea, Le et al. expand the feature vector (824 dimensions) by adding the statistical indicators (i.e. the number of words in the copied file) to obtain more detailed user behavior characteristics \cite{le2020analyzing}. However, such an expansion has the benefits of improving the information richness, but it also introduces risks associated with information redundancy and huge overhead. Unlike the above methods, Chattopadhyay et al. are not limited to simple frequency and statistics aspects, but introduce the concepts of sliding window and time-series feature to capture the dynamic characteristics of user activities \cite{chattopadhyay2018scenario}. More specifically, they construct the feature vector by calculating the variation of each indicator within the time window. Yuan et al. pay more attention to the activity time information, so they extract the behavior temporal representation from both the intra-session and inter-session levels \cite{yuan2019insider}. That is to say, they generate the origin behavior features by calculating activity times, activity types, session durations, and session intervals. In order to detect the low-signals yet long-lasting threats, Yuan et al. add the group-related indicators and behavioral deviation indicators based on the original single-day features, and construct a compound matrix to characterize the user’s historical behavior pattern \cite{yuan2021time}.

The hidden features based on representation learning is another common feature extraction method. It exploits the deep learning model to automatically extract user’s behavior characteristics. In a sense, this method can be regarded as multiple abstractions of raw audit data, and its purpose is to obtain numerical representation that is most conducive to anomaly detection. Sharma \cite{sharma2020user} and Yuan \cite{yuan2018insider} et al. arrange the activities in the audit data in chronological order for each user to generate the activity sequence set, and then send those sequences into Long Short Term Memory (LSTM) network to obtain the advanced behavior representation. Sun et al. adopt the same network model (i.e. LSTM) to capture the general nonlinear dependency over the history activities, but the difference is that they use the interleaved sequences formed by user behaviors and user-attributes as model input \cite{sun2021deepmit}. Based on the Bidirectional Encoder Representations from Transformers (BERT) model, Yuan et al. map the activity type and its corresponding time information to the embedding space, and then construct the behavior representation by summing the above vectors \cite{yuan2020few}. To further improve the model accuracy, Jiang et al. expand the feature vector by exploiting the graph convolutional network and structural information between users \cite{jiang2019anomaly}. Inspired by natural language processing, Liu et al. first use the ‘4W’ template to reorganize the audit logs, and then transform the human-consumable textual data into the machinable-consumable numerical vector with the help of Word2vec model \cite{liu2019insider}. The main advantage of this approach is that it can capture the potential semantic properties in the original audit logs without relying on any domain knowledge. Meanwhile, it suffers from the defect of limited detection performance.

\textbf{\emph{2) Unsupervised Anomaly Detection.}} Considering the feasibility in practical applications, unsupervised anomaly detection methods are the current mainstream study direction. Many classical unsupervised anomaly detection algorithms such as IF \cite{gavai2015detecting}, HMM \cite{rashid2016new}, AutoEncoder (AE) \cite{sharma2020user,liu2018anomaly,yuan2021recompose,liu2019unsupervised,le2020exploring}, Generative Adversarial Network (GAN) \cite{yuan2020data,gayathri2021anomaly} and One Class Support Vector Machine (OCSVM) \cite{lin2017insider} have been applied in the field of insider threat detection. Gavai et al. design an insider threat detection scheme based on enterprise online activity data \cite{gavai2015detecting}. The scheme does not attempt to model the normal user behavior, but utilizes the isolation forest algorithm to detect statistical outliers directly. Rashid et al. apply the Hidden Markov Model to insider threat detection for the first time \cite{rashid2016new}. They use the user behavior sequences as system input, the hidden Markov model as modeling approach, and the deviation between predicted results and actual operations as judging criteria to detect anomalies. The benefit of this method is that the detection model has good interpretation, which is convenient for experts to conduct post-event analysis. But it also has some disadvantages such as high computational overhead and poor early detection performance. Yuan et al. propose a novel LSTM-based deep autoencoder-based anomaly detection method for discrete event logs, which determines whether a sequence is normal or not by analyzing (encoding) and reconstructing (decoding) the given sequence \cite{yuan2021recompose}. Moreover, Yuan \cite{yuan2020data} and Gayathri \cite{gayathri2021anomaly} et al. utilize the generative adversarial network to augment training data, and such a solution has resulted in good effect.
The powerful representation and discrimination capabilities of deep learning model provide new opportunity for insider threat detection. Most unsupervised anomaly detection methods can be roughly categorized into two approaches: reconstruction-based anomaly score and reconstruction-based representation learning. The former identifies anomalies based on whether the construction error exceeds the threshold, and classical methods belonging to this category include AutoEncoder \cite{sharma2020user,liu2018anomaly,yuan2021recompose,liu2019unsupervised,le2020exploring} and GAN \cite{yuan2020data,gayathri2021anomaly}. The latter utilizes low-density rejection principle for the extracted hidden features to detect anomalies, and examples of such methods are OCSVM \cite{lin2017insider} and kernel density estimation (KDE) \cite{maloof2007elicit}. Although the above works have distinctive views and new ideas on basic feature extraction, their final discriminative methods are nothing more than the above two principles. For example, Liu et al. combine the word2vec model with an autoencoder classifier to detect insider threat \cite{liu2019unsupervised}, while Lin et al. achieve the same goal by exploiting deep belief network and one class support vector machine \cite{lin2017insider}.

\begin{figure*}[!htbp]
	\centering 
	% Uncomment below line to include image
	\includegraphics[width=1\textwidth]{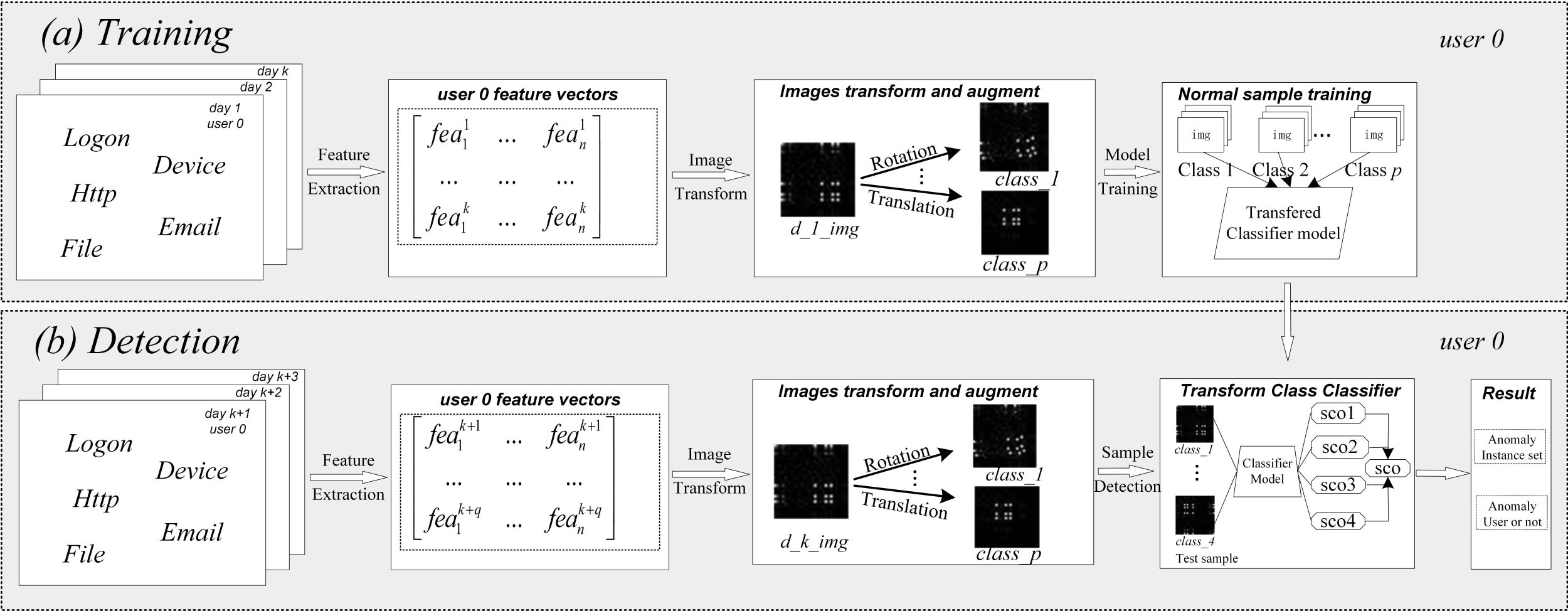}% <image name>
	\caption{\label{fig1}Overview of IGT Mechanism.}
\end{figure*}

Compared with the upper works, IGT scheme focuses on the technique innovation in the aspect of insider threat detection method, and provides an effective reference for other anomaly detection problems in the security field. In this regard, GeoTransform \cite{golan2018deep} is similar to our work, but its target is a well-arranged visual image. While the target of IGT is discrete multi-source audit log, which greatly increases the problem complexity. Moreover, the major differences between our work and the classical unsupervised scheme proposed in \cite{liu2019unsupervised} are as follows: i) IGT constructs the separate behavior model for each user, instead of sharing one model for all users, because we believe that different users have different behavior patterns, and a unified model is difficult to detect the subtle malicious activities. ii) IGT adopts the statistical feature extraction method based on expert domain knowledge, while the scheme in \cite{liu2019unsupervised} uses the Word2vec model to extract behavior characteristics. Since the malicious behavior is closely related to the context in which it occurs, the intervention of domain knowledge is conducive to improve the modeling accuracy. iii) Unlike the work \cite{liu2019unsupervised}, IGT utilizes a completely different anomaly detection method, which converts unsupervised anomaly detection problem into supervised image classification problem by constructing a self-labelled dataset, thereby reducing the problem complexity. More importantly, the performance of our proposed solution is significantly better than the existing mainstream insider threat detection methods.

\section{Methodology}
The primary purpose of this study is to explore the possibility of image-based classification methods in the field of insider threat detection, thereby providing new research idea for solving the cybersecurity problem. To this end, this section begins with the workflow of our proposed method, and summarizes the overview of design principle and basic framework (Section III-A). Then we elaborate the feature extraction and image conversion steps involved in the detection procedure respectively (Section III-B and Section III-C). Finally, the modified unsupervised anomaly detection method and corresponding algorithm implementation are presented emphatically in the Section III-D.

\subsection{System Overview}
Figure \ref{fig1}  shows the system workflow of IGT, in which there are four key procedures, the feature extraction, image conversion, anomaly detection and result analysis. The feature extraction is responsible for the abstraction and generalization of audit data. According to the potential malicious scenarios, this procedure extracts the frequency and statistical features for each user by aggregating multiple audit data, and then constructs the numerical vector representing user behavior and profile information. After obtaining the initial feature vectors, the image conversion procedure generates image creation tasks and sends these grayscale images to the anomaly detection module. On the basis of these grayscale images, the anomaly detection procedure can construct and train a geometric transformation-based classification model. This discriminative model is applied on transformed instances of the test samples, and those with poor classification accuracy are regarded as suspicious instances. When finishing the classification work, the result analysis procedure conducts the identification of malicious instances and users according the corresponding threshold, and provides the final detection results to the security analyst.

Before getting into the details of this method, we will state the insider threat detection problem studied in this paper clearly and give a simple mathematical formulation. This problem can be defined as: “Given employee’s past online activity, predict if an employee is behaving abnormally either with respect to his past activity at any given time instance.” Let $\bar{X} $  be the space of all activities, and let $X \subseteq \bar{X} $  be the set of normal activities. Given a sample $S \subseteq X $  and employee’s past activities, we would like to construct the best possible classifier $h_s(x):\bar{X}\rightarrow{\{0,1\}} $, where $h_s(x)=0 \Leftrightarrow \mathrm{x} \in X$. Activities that are not in $X$ are regarded as anomalies. A common method for controlling the classification result is to learn a scoring function $r_s(x):\bar{X} \rightarrow{R} $, such that higher scores mean that samples are not more likely to be in $X$ \cite{golan2018deep}. Once such a scoring function has been learned, a classifier can be constructed from it by specifying an anomaly threshold ($\lambda$):
\begin{equation}
h_s^\lambda(x)= \begin{cases}
1, \quad r_s(x)\geq \lambda \\
0, \quad r_s(x)< \lambda
\end{cases}
\end{equation}
In fact, the critical point of insider threat detection is how to construct the best possible classifier $h_s(x) $ and learn the scoring function $r_s(x)$. As for the threshold, it can be determined based on empirical knowledge and numerical experiments.

As previously mentioned, the starting idea of this work is to apply the novel unsupervised anomaly detection method originated from computer vision to insider threat detection problem. That is to say, how to create the proper image representation and construct the effective discriminative model are the primary problems we should solve. In this process, the content of sample  $\boldsymbol{S}$ moves from multiple logs to grayscale images, and the type of classifier moves from unsupervised to supervised. However, the supervised classifier here does not mean that we train this discriminative model with the behavior label (normal or malicious), but using self-labelled information (i.e. rotation, translation, etc.). Although we assume that all the training data are positive samples, the real anomaly labels are not used in the whole detection process. In order to present the proposed scheme detailly, we select the benchmarking datasets provided by CERT as application example. Besides, our analysis will distinguish between malicious instances detected and malicious users detected, which represent different aspects of the proposed scheme’s performance. Because the diversity of user’s role has a strong impact on the distribution of actions performed, the high malicious instance detection rate is not synonymous with all malicious users being detected \cite{le2020analyzing}. As for a detected malicious user here, he is identified if the proportion of suspicious instances exceeds another specific threshold $\kappa$.

\subsection{Feature Extraction}
Feature extraction is an essential part of insider threat detection. The performance of anomaly detection algorithm is closely relevant to the feature vector used to train the model, but it is not a trivial work to characterize the user behavior pattern due to various factors such as validity and interpretability. In order to solve this problem, we design a more reasonable feature set based on the potential malicious scenarios. The audit data in the CERT dataset consists mainly of logon information, file information handled by users, external device information, email communication information, http detailed information of browsing history and organization’s structure information. The first five logs record most activities performed by users with a certain timeframe and provide the basic data support for user behavior analysis. The organization structure information represents context data, which can be working role and personal information. Normally, it is used as auxiliary information in the process of behavior analysis.

\begin{table}[]\footnotesize
	\setlength{\tabcolsep}{1.1mm}
	\caption{\label{tbl1}Feature Structure.}
	\centering 
	\def\arraystretch{1.15} % To increase the row spacing
	\begin{tabular}{|c|c|l|l|l|l|}
		\hline
		\multicolumn{6}{|c|}{Feature = Time + PC + Activity}                                                                                          \\ \hline
		Time & \multicolumn{2}{c|}{workhour, afterhour, weekend,whole} & PC & \multicolumn{2}{c|}{own, others', whole} \\ \hline
		& Logon  & \multicolumn{4}{l|}{logon\_times}                                           \\ \cline{2-6} 
		& File   & \multicolumn{4}{l|}{file\_oper\_times, doc\_oper\_times, exe\_oper\_times}  \\ \cline{2-6} 
		& Http   & \multicolumn{4}{l|}{http\_oper\_times, job\_oper\_times, leak\_oper\_times} \\ \cline{2-6} 
		& Email  & \multicolumn{4}{l|}{email\_oper\_times, email\_mean\_size\_atts}            \\ \cline{2-6} 
		\multirow{-5}{*}{Activity} & Device & \multicolumn{4}{l|}{usb\_oper\_times}                                       \\ \hline
	\end{tabular}
\end{table}
Based on the above audit data, we can perform feature extraction to create numerical vectors that represent user behavior pattern mostly. Given an aggregation condition $t$ (i.e. timeframe), different audit data are aggregated based on user id in chronological order, and then feature extraction is performed on the aggregated data to generate fixed-length vectors $x_t$  (also called data instances). Considering that malicious activities scatter in numerous normal working activities and potential insider attack manifests in various forms, such as data leakage and intellectual property theft, we extract behavior features from three aspects of time, computer and activity. Table \ref{tbl1} depicts the feature structure in the case of CERT dataset. Since time is an important dimension to represent user behavior pattern, we set 4 different time-frames to capture the comprehensive information as much as possible. Meanwhile, in view of the fact that masqueraders make up a large chunk of insiders, the assigned computer information is added to assist characterize the user behavior. As for the activity aspect, we design several indicators for different behavior domains, such as \emph{number of logon, number of doc file operation, number of recruiting websites visited, mean size of email attachments and number of external device used}. These indicators mainly involve two types: frequency indicator and statistic indicator, in which the former is the number of activities performed in specific timeframe and the latter is the descriptive statistic such as mean, standard deviation. Collectively, occurrence time, the assigned computer and activity indicators constitute the body of final feature, and the feature vector constructed is essentially an enumeration over all the above indicators. Furthermore, IGT adopts a categorizing scheme (e.g. \emph{job} and \emph{leak} for websites, \emph{doc} and \emph{exe} for files) when extracting HTTP and file features. Such a design can avoid the privacy leakage effectively, because it only requires examining the websites domain or file suffix instead of inspecting the specific content.

To further explore the impact of feature extraction on detection performance, we also set three different aggregation granularities based on time duration. Table \ref{tbl2} shows the details of aggregation granularity in this paper. User-week and user-day data instances represent the users’ activities during the corresponding time frame. User-session data instances summarize the behavior information by collecting activities from Logon to corresponding Logoff; or from one Logon to next Logon. According to the previous feature structure, we construct 40, 28 and 16 features for different data instances, respectively. However, the finer-grained data don’t mean the best detection performance. This is because the difficulty of anomaly detection increases with the refinement of data granularity. Although the finer-grained data can provide higher data fidelity for behavior analysis, it also has the drawback of long learning time and large imbalanced ratio. Moreover, the duration and number of activities vary greatly from one session to another, which further increase the problem complexity. That is to say, there is a tradeoff between detection efficiency and data fidelity, and it is also one of our jobs in this work to explore which aggregation granularity is most beneficial for insider threat detection.

It should be noted that there are several similar feature extraction schemes in the academia \cite{chattopadhyay2018scenario,le2020analyzing}. To make it easier to understand our innovation, we make the following comparison. What is common between all these schemes is that they extract the behavior features based on experts’ domain knowledge. But compared with these previous studies, our feature extraction method is designed for malicious behavior with a higher degree of pertinence. Although the number of features proposed in work \cite{le2020analyzing} is much more than our scheme which in turn increase the information richness, it also introduces risks associated with redundancy and overhead. Due to high dependency between behavior features and the existence of noise, numerous input variables (features) sometimes degenerate the model performance and increase the unnecessary overhead. Therefore, we strip away the inessential indicators and design several critical features related to potential malicious scenarios. Similar to our work, the method proposed in work \cite{chattopadhyay2018scenario} extracts 20 specific behavior features, but they only consider two aspects of information: time and activity. Additional experiment analysis is performed in Section IV-B to verify the superiority of our feature extraction scheme.
\begin{table}[]\footnotesize
	\setlength{\tabcolsep}{1.1mm}
	\caption{\label{tbl2}Feature Granularity Detail(Sc:Based on r4.2).}
	\centering 
	\def\arraystretch{1.15}
	\begin{tabular}{|c|c|c|c|}
		\hline
		Level   & Notation & Meaning                                 & Dimension \\ \hline
		Week    & $x_w$     & User behavior data aggregated by week   & 40                \\ \hline
		Day     & $x_d$       & User behavior data aggregated by day    & 28                \\ \hline
		Session & $x_s$       & User session data, from logon to logoff & 16                \\ \hline
	\end{tabular}
\end{table}

\subsection{Image Conversion}
The extracted behavior feature vector cannot be used for unsupervised anomaly detection algorithm based on geometric transformation because of its column vector format. Thus, how to convert the origin feature vector to grayscale image is a problem worthy of exploring. Actually, the essence of image conversion is to generate the square matrix based on column vector. In this regard, we propose a heuristic image conversion strategy. Given that the malicious scenario is closely related with various features, we tend to represent all the incidence information between the features directly to enhance the expression. Thus, we use the following equation to perform the image conversion:
\begin{equation}
s = x^T\cdot x
\end{equation}

Although such a conversion doesn’t add the essential information than original version, it constructs the spatial relationship and provides a forthright representation among the features. In the earlier stages of this work, we tried a few information enhancement methods such as convolutional coder to construct more complicated images, which degraded performance and we abandoned them altogether. We hypothesize that these methods augment much irrelevant information, and then interfere the learning effect. By means of the above conversion, each feature vector extracted from the audit data can be represented as the grayscale images.

\subsection{Anomaly Detection}
After obtaining the user behavior image representation, we can apply the unsupervised anomaly detection algorithm based on transformation to detect malicious instances. The anomaly detection procedure can be divided into two major steps. Firstly, we create a self-labelled dataset of images from the origin user behavior image set $\boldsymbol{S}$, by using a series of geometric transformations $\boldsymbol{\varPsi}$. As for the reason why use geometric transformations in here, we will explain it later. Let $\boldsymbol{S}_{\boldsymbol{\varPsi}}$ denotes the created new dataset, which is generated by applying each transformation in $\boldsymbol{\varPsi}$ on all images in $\boldsymbol{S}$. The label of new dataset instance is the index of the transformation that was applied on it. That way, we generate a self-labelled multi-class dataset (with $|\Psi|$ classes) whose cardinality is $|\boldsymbol{\varPsi}||\boldsymbol{S}|$ . Then we can train a multi-class image classifier $h_s(x)$ to predict the transformation index of image. In testing phrase, this classifier is applied on transformed instances of test sample, and the sample with worse classification results will be judged as malicious. To measure the quality of classification results, we utilize a scoring function $r_s(x) $ , which is defined as the combination of the log-likelihood of the output softmax vectors $y(x)\triangleq \operatorname{soft} \max \left(h_{s}(x)\right)$  coming from the classifier $h_s(x)$. After that we select the proper threshold based on numerical experiments and report those instances whose scores exceed the threshold to analysts for further identification.

\begin{algorithm*}
	\caption{Insider Threat Detection}
	\begin{algorithmic}[1] %每行显示行号
		\Require user set $\boldsymbol{U}$, user behavior set $\boldsymbol\varOmega$, transformation set $\boldsymbol\varPsi$, softmax classifier $h_s(x)$
		\Ensure suspicious behavior set $\boldsymbol\varPhi$, suspicious user set $\boldsymbol{U_a}$
		\State $\boldsymbol\varPhi=\varnothing$, $\boldsymbol{U_a}=\varnothing$, $\boldsymbol\varGamma=\{week,day,session\}$
		\For {$u \in \boldsymbol{U}$} 
		\State $\boldsymbol{\varOmega}_{u}=\boldsymbol{\varOmega}_{\text {train }}+\boldsymbol{\varOmega}_{\text {test }}$
		\For {$\omega \in \boldsymbol{\varOmega}_{\text {train }}$}
		\For {$t \in \boldsymbol{\varGamma}$}
		\State calculate the feature vectors $x_t^u$ based on user id $u$;
		\EndFor
		\EndFor
		\State $\boldsymbol{S}_{t}^{u} \leftarrow\left\{x_{t}^{n}: t \in \Gamma\right\} \quad$   //	create the grayscale image $\boldsymbol{S}$ according to equation (2)
		\State $\boldsymbol{S}_{\boldsymbol{\varPsi}} \leftarrow\left\{\left(\varPsi_{j}(s), j\right): s \in S_{t}^{u}, \varPsi_{j} \in \boldsymbol{\varPsi}\right\} \quad$  //create the self-labelled dataset
		\While {not converged }
		\State train $h_s(x)$ on the self-labelled dataset $\boldsymbol{S}_{\boldsymbol{\varPsi}}$
		\EndWhile
		\For {$i \in\{0, \ldots, K-1\} $}      $\quad //K=|\boldsymbol{\varPsi}|$	
		\State calculate $\alpha_i$ according to the numerical method \cite{wicker2008maximum}
		\EndFor
		\State calculate the threshold parameter $\lambda$ and $\kappa$
		\For {$\omega \in \boldsymbol{\varOmega_{test}}$}
		\State create the feature vector $x_t^u$ , grayscale image $s_t^u$ and self-labelled images $\boldsymbol{S_{\boldsymbol{\varPsi}}}$
		\State calculate the anomaly score $r(x) \stackrel{\Delta}{=} 1-\frac{1}{K} \sum_{i=0}^{K-1}\left(\alpha_{i}-1\right) \cdot \log h\left(\varPsi_{i}(x)\right)$
		\If{$r(x)>\lambda$}
		\State $\boldsymbol{\varPhi_u} = \boldsymbol{\varPhi_u} \cup {{x}} $
		\EndIf
		\EndFor
		\If {$\left|\boldsymbol{\varPhi}_{u}\right|>\kappa$}
		\State $\boldsymbol{U}_{a}=\boldsymbol{U}_{a} \cup\{u\}$
		\State $\boldsymbol{\varPhi}=\boldsymbol{\varPhi} \cup \boldsymbol{\varPhi}_{u}$
		\EndIf
		\EndFor
		\State return suspicious behavior set $\boldsymbol{\varPhi}$ and suspicious user set $\boldsymbol{U_a}$
	\end{algorithmic}
\end{algorithm*}

During the above process, the selection of transformation set is critical to the anomaly detection performance since we think the appropriate set of transformations is problem dependent but is not fixed. In the field of image anomaly detection, the GeoTransform \cite{golan2018deep} method applies 72 different transformations for each sample. Its original intention is that these geometric transformations can preserve the spatial information and local pixel correlation of normal images. However, considering the fact that the behavior representation images are converted by feature vectors, the initial candidate transformations are not limited to geometric filed. Thus, non-geometric transformations such as Gaussian blurring and Laplace sharpening are also added to the candidate set to provide more possibility for detection performance. In addition, we don’t make any further division for rotation transformation. That is, the candidate transformations are composed of 9 types of shifts, rotations, flips, Gaussian blurring and Laplace sharpening of images in each sample, yield a set of 144 transformations $\boldsymbol{\varPsi}^{\prime}=\left\{\varPsi_{1}, \varPsi_{2}, \ldots, \varPsi_{K-1}\right\}$, where $K=144$. The candidate transformation can be described as follows: 
\begin{gather}
\left. \Biggl\{ \boldsymbol{\varPsi}^{\prime} = \varPsi_{o}^{\emph{rot}} \circ \varPsi_{s_{h}, s_{w}}^{\emph{trans }} \circ \varPsi_{f}^{\emph{flip}} \circ \varPsi_{b}^{\emph{blur}} \circ \varPsi_{p}^{\emph{sharp}}: \right. \notag \\
\left.\begin{array}{l}f, o, p, b \in\{T, F\} \\ s_{h}, s_{w} \in\{-1,0,-1\}\end{array}\right\}
\end{gather}
Where $ \varPsi_{o}^{\emph{rot}},\varPsi_{f}^{\emph{flip}},\varPsi_{b}^{\emph{blur}},\varPsi_{p}^{\emph{sharp}}$  denote the transformation of rotation (90 degree), horizontal flip, Gaussian blurring and Laplace sharpening respectively. The corresponding parameter $o,f,p,b$ indicate whether the transformations occur or not. The translation transformation is represented by $\varPsi_{s_{h}, s_{w}}^{\emph{trans }}$ , where $s_h$ and $s_w$ denotes the direction of the translation in each axis. For both filters a kernel size of $3\times3$ is used, for the Gaussian kernel we used $\sigma = 1$ and for Laplacian kernel $\sigma = 0.5$. Inevitably, such an extension will introduce unnecessary redundancy and an effective strategy should be designed to discard useless transformations. For this purpose, we combine the optimization technique \cite{reyes2020transformation} to select the transformations that are most helpful to improve the detection performance. Specifically, we splits the self-labelled dataset $\boldsymbol{S}_{\boldsymbol{\varPsi}^{\prime}}$  into binary subsets $\boldsymbol{S}_{\boldsymbol{\varPsi}_{ij}^{\prime}}$  composed of a pair of transformations $\varPsi_i $ and $\varPsi_j $ , where $i\neq j$ and $i>j$, and then calculates the accuracy for every pair of transformations by training a deep neural classifier based on Wide Residual Network model (see Section IV-A for detail) \cite{golan2018deep}. Those transformation subsets with the accuracy around 50\% would be regard as redundancy and in which only the transformation with the least number of operations would be reserved. The intuition behind this method is that both transformations are equivalent in detection performance if the classifier cannot distinguish one transformation from the other. By applying the aforementioned procedure to the candidate set $\boldsymbol\varPsi^{\prime}$, we can obtain the final transformation set $\boldsymbol\varPsi$ for the CERT dataset. More surprisingly, the experiment result shows that the geometric transformations get better results than non-geometric ones, and we think the possible reason is that the non-geometric transformation eliminate the features from $\boldsymbol{S}_{\boldsymbol{\varPsi}}$ that are important to characterize normal behavior pattern. The more experiment details about the selection of transformation set are presented in Section IV-C.

Besides, the scoring function is used to measure the anomaly degree of data instance and generate the suspicious sample set for the security analysts. Herein, we use the following scoring function as the discriminate criterion:
\begin{equation}
r(x) \stackrel{\Delta}{=} 1-\frac{1}{K-1} \sum_{i=0}^{K-1} \log p\left(y(\varPsi(x)) \mid \varPsi=\varPsi_{i}\right)
\end{equation}
Which is the combined log-likelihood of the output softmax vectors coming from the classifier $h_s(x)$, under the assumption that all of these conditional distributions are independent. However, this ideal assumption is inconsistent with facts and we replace it with the more common Dirichlet distribution \cite{golan2018deep}. Therefore, the final scoring function used in this paper is as follows:
\begin{equation}
r(x) \stackrel{\Delta}{=} 1-\frac{1}{K} \sum_{i=0}^{K-1} (\alpha_i -1) \cdot \log y \left(\varPsi_i (x) \right)
\end{equation}

Where $\alpha_i $ is the maximum likelihood parameter of Dirichlet distribution, and it can be estimated through numerical methods \cite{wicker2008maximum}. For each test sample, its log-likelihood is calculated by using the classifier output and the respective transformation’s Dirichlet parameter $\alpha_i $, and then all the log-likelihoods are summed up to yield the score. The larger the score, the more anomalous the sample. Moreover, in order to strike a balance between detection accuracy and investigation overhead, we set a specific threshold to discriminate which samples should be reported based on numerical experiments. Algorithm 1 shows the details of the whole insider threat detection mechanism.

\section{Evaluations}
In this section, we present the experimental evaluation of the proposed detection mechanism based on the CERT insider threat dataset. Firstly, we give a brief introduction about this dataset, the evaluation metrics and the deep neural classifier used in this paper, and then verify the effectiveness of the proposed features. Subsequently, we discuss the impact of the selection of transformation set and threshold on the algorithm performance. Finally, the performance comparison with other representative algorithms is presented in detail.

\subsection{Dataset}
In order to evaluate the performance of the proposed scheme, we use the Carnegie Mellon University (CMU) CERT insider threat dataset, which is a publicly available dataset for insider threat mitigation research \cite{glasser2013bridging}. The dataset consists of various versions, and each release characterizes an organization with 1000 to 4000 employees. In this paper, we select a release version r4.2 to perform evaluation. The dataset records the user activity logs of different organizations over the period of 18 months. As the insider threat events are usually rare in the real world, the class imbalance problem is also embodied fully in these datasets. For example, the number of malicious users is around a tenth of normal users in the r4.2 dataset, and the imbalance ratio of instances is even bigger. More details about the datasets can be seen in Table \ref{tbl3}. As shown in the table, the imbalance ratio is increased with the refinement of aggregation granularity, which further indicates that finer-grained data doesn’t mean the higher performance for insider threat detection. Moreover, the dataset is split into a training set and a testing set in chronological order, and the splitting ratio is set to 30\% as recommended in work \cite{chattopadhyay2018scenario}.

\begin{table}[]\footnotesize
	\setlength{\tabcolsep}{1.1mm}
	\caption{\label{tbl3}Summary of Dataset.}
	\centering 
	\def\arraystretch{1.15}
	\begin{tabular}{|c|c|c|c|}
		\hline
		Dataset               & Mal\_user:Nor\_user     & Granularity & Mal\_instances:Nor\_instances \\ \hline
		\multirow{3}{*}{r4.2} & \multirow{3}{*}{70:930} & Week        & 316:66850                     \\ \cline{3-4} 
		&                         & Day         & 966:329486                    \\ \cline{3-4} 
		&                         & Session     & 1114:469497                   \\ \hline
	\end{tabular}
\end{table}

\begin{table*}[] \footnotesize
	\setlength{\tabcolsep}{3mm}
	\caption{\label{tbl4}Instanced-based Results: Our feature VS other works.}
	\centering 
	\def\arraystretch{1.15}
	\begin{tabular}{|c|l|l|l|l|l|l|l|l|l|l|l|l|l|}
		\hline
		\multicolumn{2}{|c|}{\multirow{3}{*}{Data type}} &
		\multicolumn{4}{c|}{Supervised} &
		\multicolumn{8}{c|}{Unsupervised} \\ \cline{3-14} 
		\multicolumn{2}{|c|}{} &
		\multicolumn{4}{c|}{Random   Forest} &
		\multicolumn{4}{c|}{Isolation   Forest} &
		\multicolumn{4}{c|}{Deep Autoencoder} \\ \cline{3-14} 
		\multicolumn{2}{|c|}{} &
		\multicolumn{1}{c|}{IPR} &
		\multicolumn{1}{c|}{IDR} &
		\multicolumn{1}{c|}{IAUC} &
		\multicolumn{1}{c|}{IFPR} &
		\multicolumn{1}{c|}{IPR} &
		\multicolumn{1}{c|}{IDR} &
		\multicolumn{1}{c|}{IAUC} &
		\multicolumn{1}{c|}{IFPR} &
		\multicolumn{1}{c|}{IPR} &
		\multicolumn{1}{c|}{IDR} &
		IAUC &
		IFPR \\ \hline
		\multirow{3}{*}{Our work} &
		$X_w$ &
		\multicolumn{1}{c|}{88.92} &
		\multicolumn{1}{c|}{50.13} &
		\multicolumn{1}{c|}{92.09} &
		\textbf{0.15} &
		\textbf{15.41} &
		\textbf{38.57} &
		\textbf{75.20} &
		\textbf{5.26} &
		39.86 &
		68.72 &
		83.65 &
		5.38 \\ \cline{2-14} 
		&
		$X_d$ &
		\textbf{91.08} &
		\textbf{59.28} &
		\textbf{94.76} &
		0.17 &
		12.86 &
		36.09 &
		73.84 &
		7.32 &
		\textbf{42.76} &
		\textbf{70.82} &
		\textbf{84.71} &
		\textbf{4.86} \\ \cline{2-14} 
		&
		$X_s$ &
		82.04 &
		41.99 &
		87.99 &
		0.22 &
		12.54 &
		30.39 &
		70.39 &
		9.34 &
		41.68 &
		69.21 &
		80.73 &
		6.81 \\ \hline
		\multirow{3}{*}{Work {[}19{]}} &
		$X_w$ &
		80.23 &
		54.68 &
		89.23 &
		0.17 &
		12.62 &
		36.31 &
		64.24 &
		8.98 &
		38.18 &
		40.51 &
		80.17 &
		7.42 \\ \cline{2-14} 
		&
		$X_d$ &
		86.01 &
		59.32 &
		92.36 &
		0.17 &
		12.04 &
		34.30 &
		63.63 &
		10.96 &
		40.74 &
		53.42 &
		83.74 &
		5.31 \\ \cline{2-14} 
		&
		$X_s$ &
		75.91 &
		40.65 &
		84.67 &
		0.23 &
		11.09 &
		25.66 &
		60.24 &
		12.90 &
		31.75 &
		47.57 &
		77.63 &
		9.26 \\ \hline
		\multicolumn{1}{|l|}{\multirow{3}{*}{Work {[}10{]}}} &
		$X_w$ &
		83.26 &
		45.43 &
		90.27 &
		0.18 &
		14.65 &
		30.26 &
		66.59 &
		5.64 &
		37.64 &
		40.48 &
		81.34 &
		5.53 \\ \cline{2-14} 
		\multicolumn{1}{|l|}{} &
		$X_d$ &
		90.23 &
		54.15 &
		91.46 &
		0.18 &
		13.48 &
		33.54 &
		71.67 &
		8.37 &
		41.25 &
		53.76 &
		84.26 &
		5.29 \\ \cline{2-14} 
		\multicolumn{1}{|l|}{} &
		$X_s$ &
		72.24 &
		42.36 &
		85.26 &
		0.24 &
		12.06 &
		25.42 &
		70.84 &
		9.68 &
		38.46 &
		41.45 &
		79.42 &
		8.84 \\ \hline
	\end{tabular}
\end{table*}

Next, we introduce the performance metric used in this paper. Detection rate (DR), Precision (PR), F1 score and False positive rate (FPR) are the commonly used in the classification field, and these metrics can equally well apply to the insider threat detection \cite{yuan2021time}. In this paper, true (false) positive (TP/FP) represents the number of malicious (normal) samples that are correctly recognized as “malicious”, and false (true) negative (FN/TN) denotes the number of malicious (normal) samples that are incorrectly recognized as “normal”. Among these metrics, Precision represents the percentage of malicious warnings generated by the system that are true, and F1-score summarizes both DR and PR as a harmonic mean. Due to the extremely skewed data, the AUROC is another important indicator to evaluate the detection performance. Basically, the larger the value of AUROC, the better the anomaly detection method is. For the sake of convenience, all the following metrics are reported in percent. In addition, as mentioned in Section III-A, we report the system performance from the respective of data instance (instance-based) and organizational user (user-based). For user-based results, a user is classified as ‘malicious’ if the number of anomaly instances in specific timeframe exceeds the specific threshold. Therefore, there are two kinds of the performance metrics in this work: Instance based (IDR, IFPR, IPr, IF1) and User based (UDR, UFPR, UPr, UF1).

\begin{figure}[!htbp]
	\centering 
	% Uncomment below line to include image
	\includegraphics[width=0.35\textwidth]{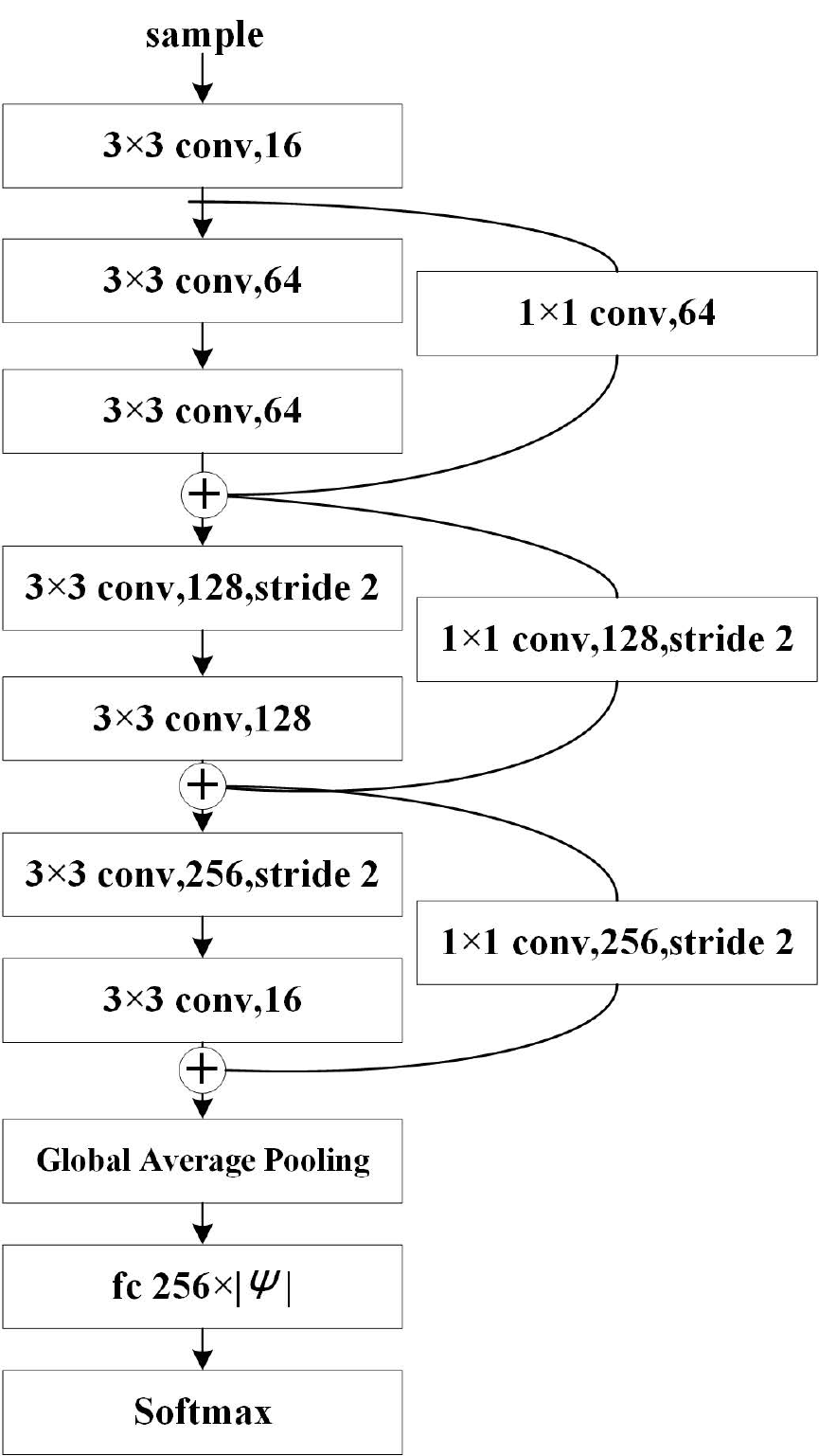}% <image name>
	\caption{\label{fig2}Wide Residual Network architecture.}
\end{figure}

\begin{equation}
DR=\frac{TP}{TP+FN}
\end{equation}
\begin{equation}
PR=\frac{TP}{TP+FP}
\end{equation}
\begin{equation}
FPR=\frac{FP}{TN+FP}
\end{equation}
\begin{equation}
F1=\frac{2}{PR^{-1}+DR^{-1}}
\end{equation}

In addition, the detail information about deep neural classifier used in the paper is presented as follows. We use a Wide Residual Network \cite{golan2018deep} with architecture parameters of depth 10 and widen factor 4 to construct the multi-class classifier $h_s(x)$. Figure \ref{fig2} depicts the full architecture of Wide Residual Network model. This model consists of 7 convolutional layers, 3 skip connections, a global average pooling and a fully connected layer with output size equal to the number of applied transformations $|\boldsymbol\varPsi|$. A batch size of 32, epoch size of 200, ADAM optimizer with defaults hyperparameters and a cross-entropy loss are applied.

\begin{table*}[] \footnotesize
	\setlength{\tabcolsep}{4.5mm}
	\caption{\label{tbl5}Transformation Selection Results.}
	\centering 
	\def\arraystretch{1.15}
	\begin{tabular}{|c|c|c|c|c|c|c|l|}
		\hline
		\multirow{3}{*}{Model} & \multirow{3}{*}{Transformation selection setup} & \multicolumn{6}{c|}{R4.2 dataset}             \\ \cline{3-8} 
		&                                      & \multicolumn{3}{c|}{AUROC (Instance)} & \multicolumn{3}{c|}{AUROC (User)} \\ \cline{3-8} 
		&                                      & $X_w$          & $X_d$         & $X_s$         & $X_w$          & $X_d$         & $X_s$        \\ \hline
		Transform36 & None   (original transformation set) & 85.34       & 87.26      & 80.69      & 89.76     & 90.69     & 88.03     \\ \hline
		Transform63 & Add   non-geometric transformations  & 85.23       & 86.94      & 79.48      & 89.01     & 89.34     & 87.26     \\ \hline
		Transform18            & Optimization selection over Transform63        & \textbf{86.31} & \textbf{88.23} & \textbf{81.47} & \textbf{90.15} & \textbf{92.13} & \textbf{89.71} \\ \hline
	\end{tabular}
\end{table*}

\subsection{Feature Superiority Comparison}
Features play an important role in determining the performance of anomaly detection. In this section, we give a contrastive analysis on the superiority of the feature extraction scheme. The comparison objects are the feature extraction methods proposed in \cite{chattopadhyay2018scenario} and \cite{le2020analyzing}. All these methods extract behavior features based on experts’ domain knowledge, but the number of features proposed in work \cite{le2020analyzing} are much more than ours, which means more detailed information about the user activities. Different from our scheme, the method proposed in work \cite{chattopadhyay2018scenario} extracts features from the aspect of time and activity. To evaluate the superiority more reasonably, we select three classical classification algorithms as the anomaly detection approach: random forest (RF) \cite{breiman2001random}, isolation forest (IF) \cite{liu2008isolation} and autoencoder \cite{liu2019unsupervised}. These approaches utilize the extracted features as input directly, and output the performance metric as the evaluation criteria. In this process, Python 3.7 is used for feature extraction and Scikit-learn is used for implementing anomaly detection algorithms. The features are normalized before being used to train the classifier. In terms of parameter selection, we perform parameter search using hyperopt, which is a parameter tuning solution based on the Parzen estimator \cite{bergstra2013making}. Specifically, for RF, we tune the number of features (all, square root and log base-2 of all features), the number of decision tree estimators (50 to 100) and the depth of individual trees (3 to 10). Similarly, for the autoencoder, a limit of 200 epochs was assumed. The number of hidden layers was searched between 1 to 3 and each hidden layer has the size set to a half of the previous layer. The mini-batch size is set to 32 and L2 regularization penalty is $10^{-6}$. With respect to the isolation forest, we tune the number of trees (30 to 100) and the threshold for suspicious instance is set 5\%. It should be noted that the isolation forest is trained on the train set and is then applied to the test set, instead of the whole dataset at a time. All the results are obtained by averaging the multiple experiment data, where each setting is randomly repeated 20 times.

\begin{figure}[!htbp]
	\centering 
	% Uncomment below line to include image
	\includegraphics[width=0.42\textwidth]{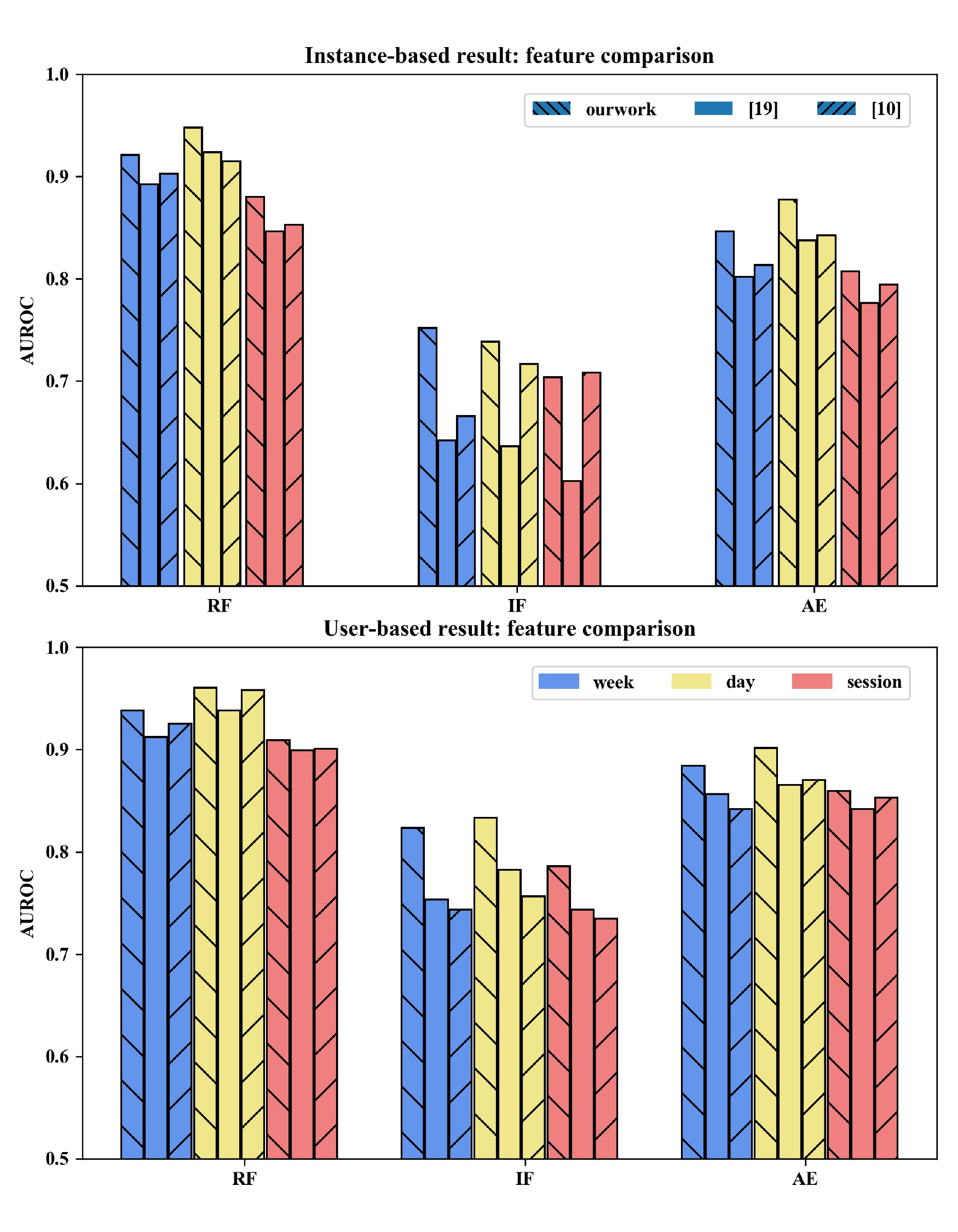}% <image name>
	\caption{\label{fig3}Feature Comparison Results.}
\end{figure}

Table \ref{tbl4} and Figure \ref{fig3} present the feature comparison results by the anomaly detection approaches on different data granularity levels. It is observed that our feature extraction scheme performs best among three schemes in any granularity level. And this advantage becomes more apparent when the detection methods change from supervised to unsupervised. The main reason is that the ground-truth information used in the supervised algorithm makes up for the deficiency of features in behavior characterization. Meanwhile, compared with the more detailed feature extraction scheme \cite{chattopadhyay2018scenario}, the more succinct method proposed in work \cite{le2020analyzing} performs better as well. This is because the numerous features (824 dimensions) inevitably introduce much redundancy, which in turn degenerate the model performance. On the other hand, the more detailed and pertinent information is beneficial to improve the detection performance when the dependency between features is nonexistent. As shown in Figure \ref{fig3}, the AUROC value of our feature extraction scheme is higher than others in both cases (instance-based and user-based results), which shows its superiority in representing user behavior pattern.
\subsection{Transformations Selection and Parameter Analysis}
As mentioned in Section III-D, we attempt to explore more possibility for detection performance by using various transformations. To obtain the proper transformation set, we apply the transformation selection procedure to the candidate set based on the r4.2 dataset. Due to the huge computational overhead brought by numerous transformations, we use a simplified setup by conducting the non-geometric transformation only over shift operations, which generates 27 possible states. That is to say, the number of candidate transformations changes from 144 to 63. Following this procedure, we compare the performance of different transformation pairs and the results are summarized in Table \ref{tbl5}.

From Table \ref{tbl5} we can see that the non-geometric transformations are not helpful to improve the detection performance. We think the possible reason is that the non-geometric transformations reduce the representational capacity of the local pixels converted by discrete feature vectors. For example, the malicious scenarios are closely related with specific features, but the non-geometric transformations such as smooth processing weaken this incidence relation. In addition, only the 18 geometric transformations (9 shift and rotation) remained after applying the optimization selection technique, but their AUROC values (Transform18) are higher than others. This phenomenon indicates that eliminating redundant transformations is beneficial to reduce computational overhead, make the algorithm faster, and lower the complexity of the classification space. Summarily, the final transformations used in the work are the translation and rotation.

\begin{figure}[!htbp]
	\centering 
	% Uncomment below line to include image
	\includegraphics[width=0.42\textwidth]{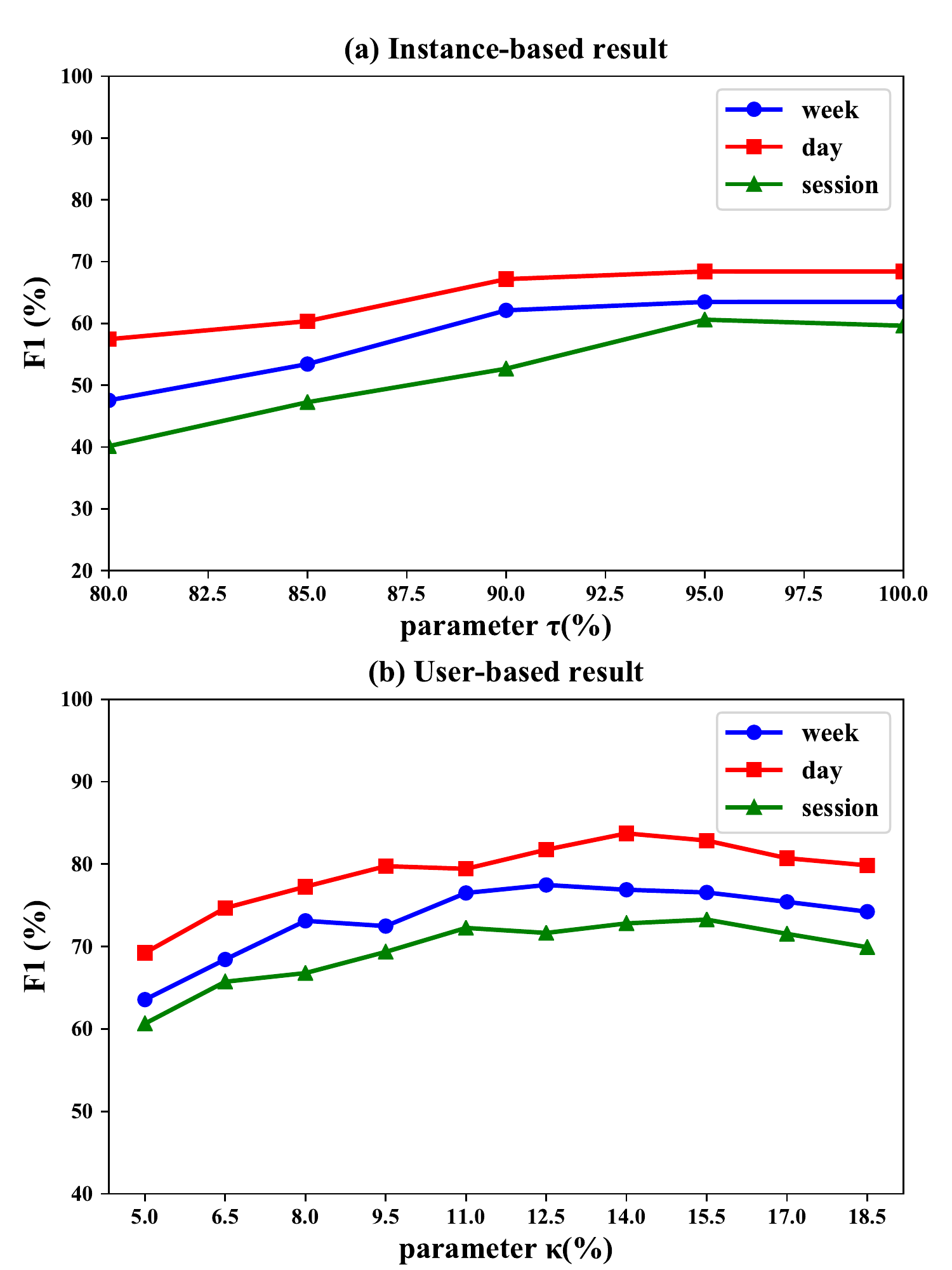}% <image name>
	\caption{\label{fig4}Relationship between threshold and performance. (a) the impact of threshold $\tau$ (i.e.$\lambda$) on the instance-based result; (b) the impact of threshold $\kappa$ on the user-based result.}
\end{figure}

When describing the IDIGE mechanism, we introduce threshold parameters $\lambda$ and $\kappa$ to discriminate the suspicious samples. The instance whose anomaly score exceeds the threshold $\lambda$ and the user whose quantity of suspicious instances exceeds $\kappa$ are regarded as anomalies and reported to the security experts for further investigation. Generally speaking, the threshold is correlated with the available investigation budget and is set primarily based on empirical knowledge and numerical experiments. However, considering that the user behavior are streaming data, we tend to select a historical score of training sample as the threshold $\lambda$ rather than a proportion of testing samples. This is, the threshold is fixed within a specified period of time, and such a setting is convenient for the subsequent discrimination of new coming behavior data. As recommended in most works, the $\tau$-percentile of historical scores is specified as the threshold. In this case, the meaning of parameter $\tau$ is synonymous with the threshold $\lambda$, just they show themselves in different forms. To assign the proper value to the parameter $\tau$, we conduct the following numerical experiments and adopt F1 performance metric for its expressive power. We analyse the impact of threshold on the algorithm performance when applying the Transform18 as the transformation set. As shown in Figure \ref{fig4}(a), for all granularity levels, the F1 metric generally shows an upward trend with the increase of threshold, but this trend pretty disappears when the threshold is close to the theoretical maximum. A larger threshold means the less alarms, which could result in many suspicious events undetected. However, in the security domain, it is preferable for managers to reduce the false negative than the false positive. Therefore, we select 95-percentile of historical anomaly scores in the training phrase as the threshold $\lambda$.

Similarly, we experimentally determine the appropriate value for threshold $\kappa$. Although the determination of malicious user depends on whether the person perform a malicious act or not in the practical situation, there is a proportional relation between the quantify of suspicious instances and anomaly degree. The more suspicious instances, the greater the likelihood that the user is malicious. But it is not reasonable to set a fixed constant for all the users because the number of malicious behaviors varies with the user’s behavioral habit and ultimate purpose. Therefore, we adopt the similar method to select the parameter $\kappa$. Specifically, we calculate the proportion of suspicious instances in latest timeframe whose length is equal to the timeframe of train dataset, and label the user as malicious if this proportion exceeds the threshold $\kappa$. The intuition behind this method is that the user with larger abnormal proportions than historical normal timeframe is more likely to be the malicious user. We test ten values for $\kappa$ ranging from 5\% to 20\% with a step size of 1.5\%. Figure \ref{fig4}(b) presents the corresponding result. It can be seen that the F1 metric of user-based detection results will decrease accordingly when the threshold $\kappa$ is set too large or too small. This is consistent with the intuition that a small $\kappa$ means more false alarms while a large $\kappa$ means more false negatives. In either case, it results in the performance degradation. Besides, from the Figure \ref{fig4}(b) we can also see that the detection performance is at the relatively high level when $\kappa$ is set to be 14\%.

\subsection{Comparison with Other Related Work}
In this section, we compare the performance of the IGT algorithm with other classical unsupervised insider threat detection algorithms, namely, autoencoder-based method \cite{liu2019unsupervised} and IF-based method \cite{gavai2015detecting}. The main differences between the work \cite{liu2019unsupervised} and our scheme has been introduced in Section II, and here we refer to their model as the baseline model. The main idea of autoencoder model is to learn the normal behavior pattern based on the construction errors. After trained with only normal samples, it can construct the normal samples with minimal reconstruction errors, and a high reconstruction error means that significant deviation has occurred between the test sample and normal samples. IF presumes that the outliers are easier to isolate from the rest of the data than normal samples, hence the samples with shorter path length to the corresponding leaves are considered suspicious. All three algorithms are trained in an unsupervised manner and label the instances whose anomaly scores exceed the 95-percentile of historical scores in the training phrase as the suspicious instances. Different from the IGT algorithm, IF and autoencoder only use the original feature vector as the inputs and don’t involve the transformation operations. In terms of experimental setting, we implement the IGT classifier with Pytorch, and the Adam optimizer with default hyperparameters is used to minimizing the cross-entropy loss function. Batch size and epochs for all these methods are set to 32 and 200. As for the IF and autoencoder models, we adopt the same experimental setting described in Section IV-B and search the best hyper-parameter by means of hyperopt tool.

\begin{figure}[!htbp]
	\centering 
	% Uncomment below line to include image
	\includegraphics[width=0.42\textwidth]{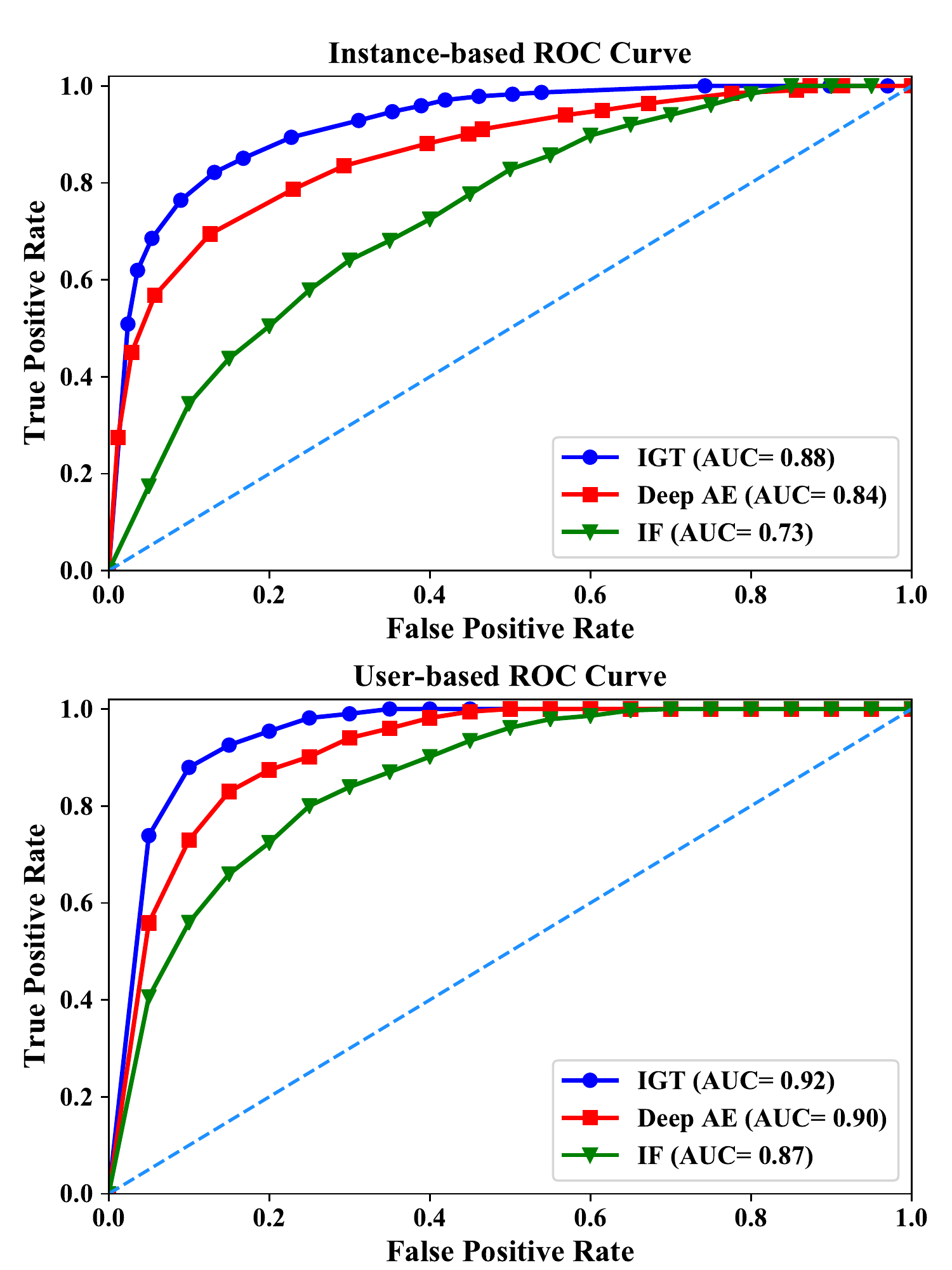}% <image name>
	\caption{\label{fig5}ROC curve under different approaches (day).}
\end{figure}
\begin{figure*}[!htbp]
	\centering 
	% Uncomment below line to include image
	\includegraphics[width=1\textwidth]{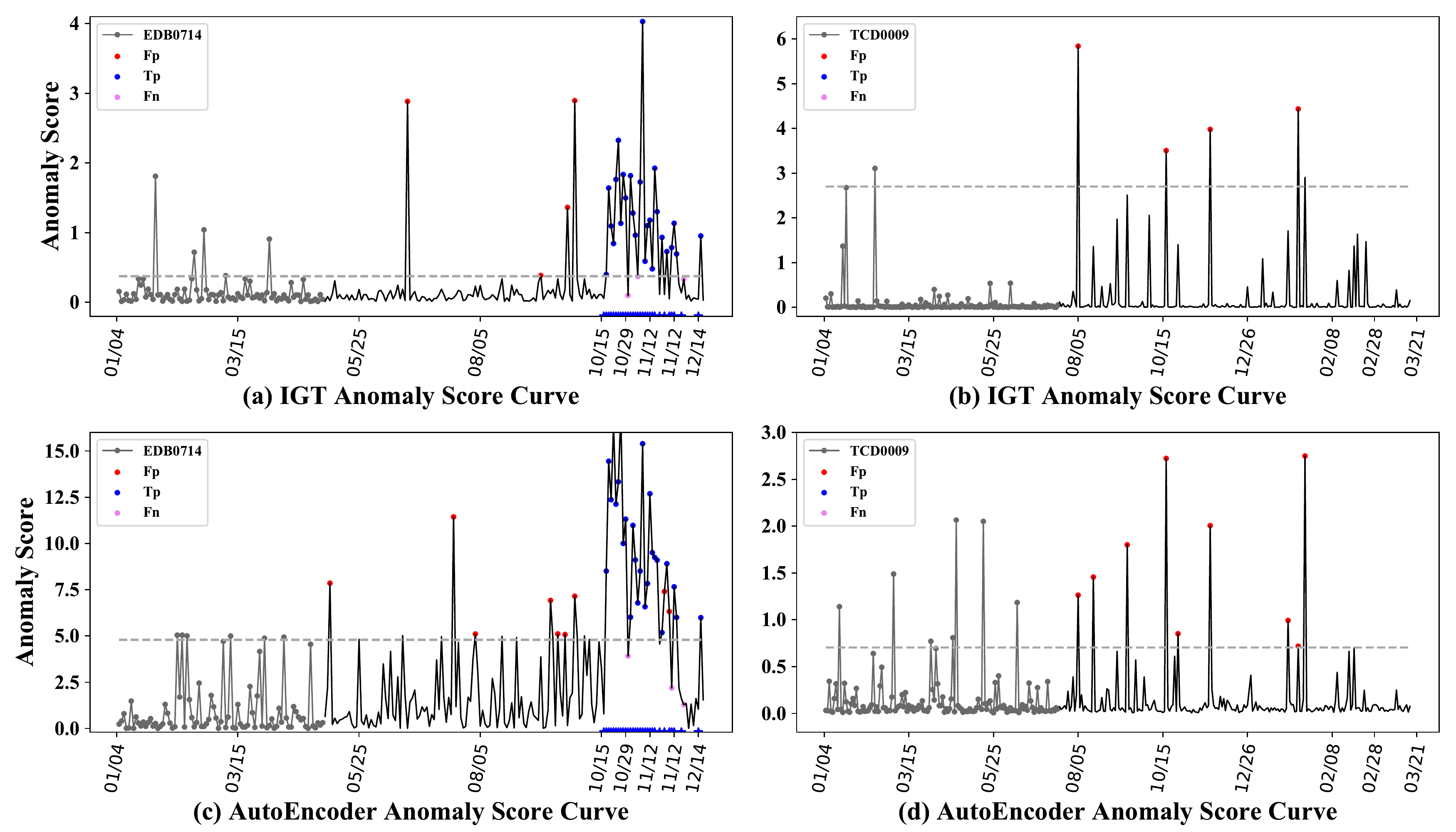}% <image name>
	\caption{\label{fig6}Trends of anomaly scores of different users (‘mal’ or ‘nor’) in r4.2 under different approaches. (a) IGT in malicious user EDB0714; (b) IGT in normal user TCD0009; (c) AutoEncoder in malicious user EDB0714; (d) AutoEncoder in normal user TCD0009.}
\end{figure*}
Firstly, we compare the performance of three different detection algorithms from two aspects, instance-based results and user-based results. Table \ref{tbl6} gives the detailed results and Figure \ref{fig5} depicts the corresponding ROC curves and AUCs on user-day data. It can be seen that the IGT performs best among these insider threat detection algorithms in either case, and the performance of AutoEncoder is better than that of IF. Compared to the precision metric (40\%-60\%), the detection rate of each algorithm is relatively high-level (60\%-80\%), which means that the critical point of improving detection performance is to reduce the false alarms. In this regard, IGT achieves the lower FPR with the similar DR. Note that the isolation forest algorithm shows a surprisingly poor performance with higher FPR and lower DR. We think the possible reason is that IF has the weaker representational learning ability and is more suitable for outlier detection rather than novelty detection. Figure \ref{fig5} further demonstrates the superiority of  IGT scheme, where the instance and user based AUROC are improved by 4\% and 2\% than AutoEncoder algorithm, respectively. Although the performance gaps between these algorithms become smaller when the object detection changes from anomaly instance to anomaly user, this is in line with our expectation and it can be explained through the smaller imbalance ratio.

Secondly, in order to explore the possible explanation for the performance difference between these algorithms, we analyse the trends of anomaly scores of two different users under different approaches based on user-day data, in which EDB0714 is malicious user and TCD0009 is normal user. The detail information can be seen in Figure \ref{fig6}. Due to the poor performance, the IF-based algorithm is ignored here. The grey line denotes the anomaly scores of training samples, and the black line denotes the anomaly scores of test samples. The false positive, true positive and false negative are depicted by red, blue and violet points, respectively. The star markers at the bottom indicates the actual anomaly days. From Figure \ref{fig6}(a) and Figure \ref{fig6}(c), we can observe that the most anomaly instances of malicious user can be detected by the IGT and AutoEncoder algorithms, but the number of red points (false positive) under AutoEncoder approach is higher than IGT. This phenomenon is more obvious in the trends of anomaly scores of normal users (i.e. Figure \ref{fig6}(b) and Figure \ref{fig6}(d)), which further indicates that the representational learning ability of AutoEncoder is weaker than IGT scheme. Although the results of two individual users may be hard enough convincing, it is needed to explain that both users are selected randomly in the organizational employee and the similar conclusion can be drawn from other users.

\begin{figure}[!htbp]
	\centering 
	% Uncomment below line to include image
	\includegraphics[width=0.42\textwidth]{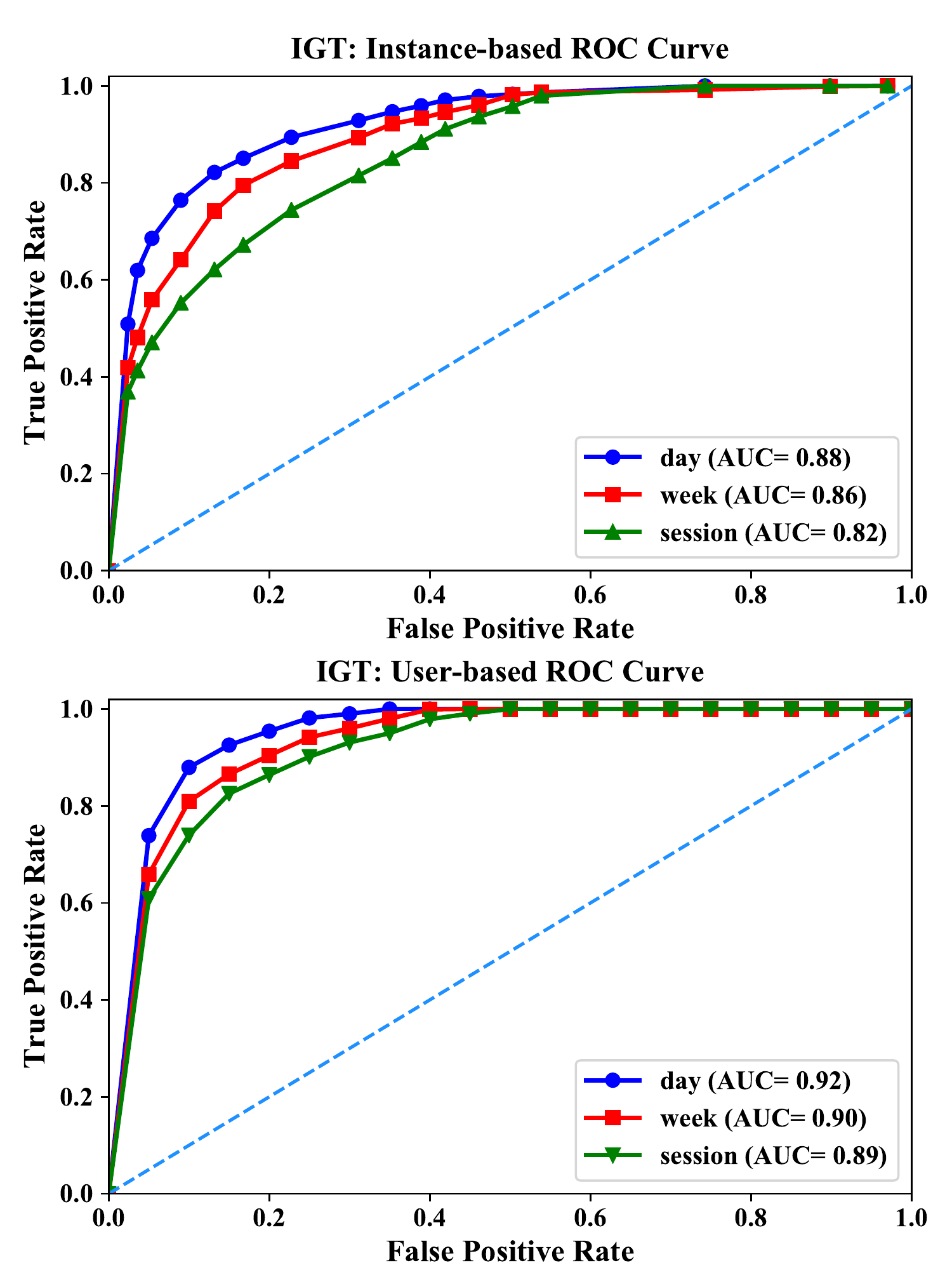}% <image name>
	\caption{\label{fig7}IGT performance on different data granularity levels.}
\end{figure}

\begin{table*}[]\footnotesize
	\setlength{\tabcolsep}{2.8mm}
	\caption{\label{tbl6}Algorithm comparison results.}
	\centering 
	\def\arraystretch{1.15}
	\begin{tabular}{|c|c|c|c|c|c|c|c|c|c|c|c|c|c|}
		\hline
		\multicolumn{2}{|c|}{Approach}       & \multicolumn{4}{c|}{IGT}     & \multicolumn{4}{c|}{Isolation Forest} & \multicolumn{4}{c|}{Deep Autoencoder} \\ \hline
		object & Level & PR    & DR    & AUC   & FPR  & PR    & DR    & AUC   & FPR  & PR    & DR    & AUC   & FPR  \\ \hline
		\multirow{3}{*}{Instance-based} & $X_w$ & 56.47 & 72.47 & 86.31 & \textbf{3.62} & 15.41    & 38.57   & 75.20   & 5.26   & 39.86    & 68.72   & 83.65   & 5.38   \\ \cline{2-14} 
		& $X_d$    & \textbf{60.18} & \textbf{74.59} & \textbf{88.23} & 4.12 & 12.86 & 36.09 & 73.84 & 7.32 & 42.76 & 70.82 & 84.71 & 4.86 \\ \cline{2-14} 
		& $X_s$    & 50.73 & 64.87 & 82.47 & 5.72 & 12.54 & 30.39 & 70.39 & 9.34 & 41.68 & 69.21 & 80.73 & 6.81 \\ \hline
		\multirow{3}{*}{User-based}     & $X_w$ & 70.29 & \textbf{84.51} & 90.15 & 0.44 & 48.76    & 56.43   & 82.34   & 0.66   & 67.23    & 78.67   & 88.42   & 0.63   \\ \cline{2-14} 
		& $X_d$    & \textbf{74.26} & 82.64 & \textbf{92.13} & 0.39 & 45.68 & 63.32 & 87.34 & 0.60 & 68.78 & 80.78 & 90.17 & 0.50 \\ \cline{2-14} 
		& $X_s$    & 65.37 & 70.63 & 89.71 & 0.54 & 32.29 & 48.21 & 78.61 & 0.89 & 61.59 & 69.25 & 85.95 & 0.62 \\ \hline
	\end{tabular}
\end{table*}
Moreover, we investigate the impact of different data granularity levels on IGT detection algorithm and the experiment results are presented in Table \ref{tbl6} and Figure \ref{fig7}. It can be seen that the relationship between the algorithm’s performance and data granularity is not monotonic. For example, the AUROC value of user-week data (0.86) is higher than that of user-session (0.82), but lower than that of user-day (0.88). This phenomenon demonstrates our previous guess that the finer-grained data don’t mean the best detection performance. Although the higher data fidelity is beneficial to construct more precise user individual behavior model, it also introduces the drawback of large imbalance ratio. Considering that the timeliness of practical anomaly detection system and the upper experiment results, we think the user-day data is a relatively proper granularity level to conduct the user behavior analysis. As for the finer-grained detection requirement, it could be satisfied by further analysis and detection after obtaining the suspicious days. In addition, what calls for special attention is that our experiment conclusion is a little different from the work \cite{chattopadhyay2018scenario} which declares that the algorithms’ performances are degrading by higher data granularity levels (i.e. monotone decreasing). This contradiction can be explained through different model construction methods, where IGT constructs the separate behavior model for each user but the latter shares one model for all users. In other words, the less training instances degrade the detection performance of IGT on user-week data.

\begin{figure}[!htbp]
	\centering 
	% Uncomment below line to include image
	\includegraphics[width=0.42\textwidth]{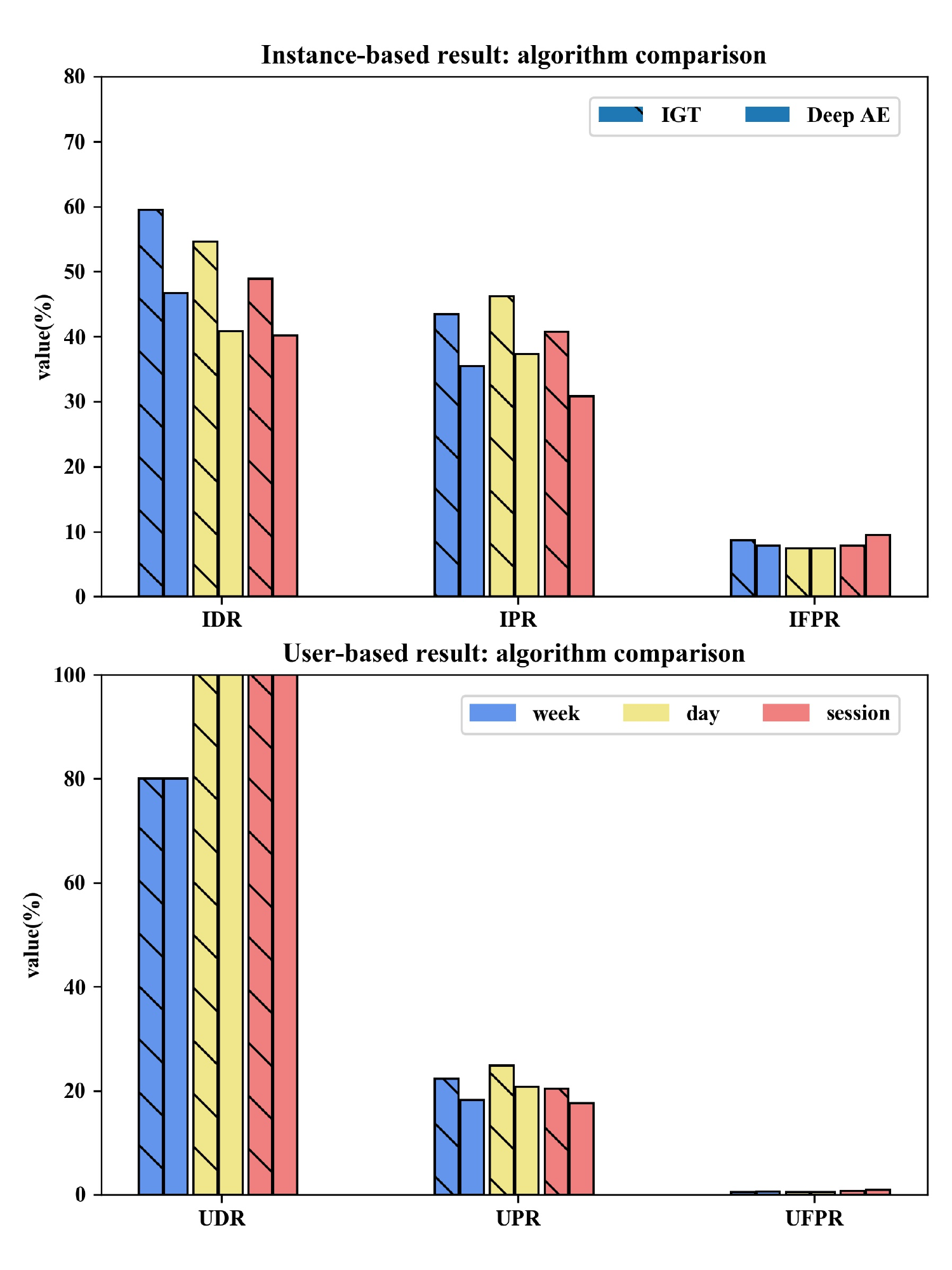}% <image name>
	\caption{\label{fig8}Algorithm Performance Comparison on r6.2.}
\end{figure}
Finally, in order to further demonstrate the effectiveness of IGT scheme, we train and test the insider threat detection algorithm based on another CERT dataset release r6.2. Compared with the r4.2 dataset, r6.2 dataset has a different organizational structure and more users (4000). Meanwhile, it simulates only one malicious user per insider threat scenario, significantly increasing the detection complexity. We adopt the same experiment setting as the preceding dataset and the experiment results are presented in Figure \ref{fig8}. As shown in the figure, although there are about 10-15\% degradation in the IDR and IPR metric, the performance of IGT is always better than the classical AutoEncoder algorithm. In response to this, we attribute the performance degradation to the larger imbalance ratio and more complex malicious scenario. In addition, since there are only 5 malicious users in this dataset, both two algorithms can detect all the malicious users, but the UFPR of IGT is significantly lower than that of AutoEncoder. On data granularity, user-day data shows higher performance than the other data types from r6.2, which further validates the previous conclusion. Overall, the IGT algorithm proposed in this paper shows better performance than the existing mainstream insider threat detection methods.

\section{ Discussion and Future Work}
\subsection{Discussion}
The features proposed in this paper are composed of occurrence time, assigned computer and specific activity, which are evaluated to have a good performance in detecting insider threat on CERT dataset. However, since feature selection is domain-specific, there is no better way to extract features that can cover all domains. Although our feature structure can cover most critical indicators, the specific activities are supposed to adjust according to the potential malicious scenarios. In other words, if the related activity information about an emerging insider threat is not contained in the feature vectors, the IGT may not be able to identify the compromise. Therefore, in order to improve the practicality and universality of the IGT mechanism, we discuss several potential concerns and measures. Firstly, which behaviors should be selected in features for a specific domain? Secondly, for specific user behavior, which attributes should be calculated as the feature representation? Thirdly, how to design the proper features to detect the unknown insider threat? In response to this, the guideline proposed by MITRE ATT\&CK provides a possible solution for the first problem \cite{Mitreattack}. Security analysts can lookup relevant behaviors for each cyber threat by means of the guideline. Next, statistical indicators and frequency indicators are the commonly used in the insider threat detection, and we should select the proper attributes according to the type of user behavior. For example, it is proper for logon activity to use the frequency indicators, while the statistical indicators such as mean and standard deviation are more suitable to content-related activities. Of course, it is also recommended to use other types of indicators to characterize user behavior pattern. At last, in terms of unknown threat detection, one possible solution is to extract the advanced features without relying on expert knowledge. Such a feature extraction method doesn’t require to make assumptions about the potential malicious scenarios, instead it adopts a purely data-driven manner to construct user behavior model, so it can provide some advantages in detecting unknown threats.
\subsection{Future Work}
In prior presentations, we introduce the basic idea and operational process of IGT mechanism detailly, and conduct numerous numerical experiments to validate its feasibility and superiority. Nevertheless, we do not provide the rigorous theoretical demonstration about the detection mechanism, as we only want to explore the possibility of image-based classification methods in the field of insider threat detection and provide a new research idea for solving the cybersecurity problem. Heuristic methods and experimental verification are the main research ideas in this paper. In addition, the behavior model used in this work is constructed based on a state description limited to a single exemplar. Therefore, there are several improvable things and the future work includes the following three directions. First, it is important to develop a theory that grounds the use of geometric transformations, which is the base of widespread application for this anomaly detection method. Second, an attempt will be made to extract the advanced behavior features without relying on expert knowledge. For example, we can obtain the potential semantic properties from the log text by means of natural language processing technique. Third, we will investigate how to utilize the temporal information in user activities to improve the detection performance.

\section{Conclusion}
Traditional insider threat detection approaches usually have the problems of low precision and high computational complexity. To solve this, in this paper, inspired by image classification technique, we propose a novel image-based insider threat detector via geometric transformation IGT. IGT construct individual behavior model for each user, applied the unsupervised classification to the images which are transformed from user behavior feature vector and further processed with geometric transformation. Unlike classical unsupervised methods, our anomaly detection approach completely alleviates the need for a generative component by converting the unsupervised anomaly detection problem into supervised image classification problem. More importantly, evaluation results on CERT dataset show that compared to classical insider threat detection approach, IGT improves the instance and user based AUROC by 4\% and 2\%. As a main future work, we will attempt to develop a theory that grounds the use of geometric transformations and explore the utilization of temporal information in user activities to further improve insider threat detection performance.

%\appendices
%\section{Proof of the First Zonklar Equation}
%Appendix one text goes here.

%\section{}
%Appendix two text goes here.

% use section* for acknowledgment
\section*{Acknowledgment}

The authors would like to thank for the discussion with Dr. Duc C. Le. This research was supported by a research grant from the National Science Foundation of China under Grant No. 61772271, and the Natural Science Foundation of Jiangsu under Grant No. SBK2020043435.

% Can use something like this to put references on a page
% by themselves when using endfloat and the captionsoff option.
\ifCLASSOPTIONcaptionsoff
  \newpage
\fi

%\end{thebibliography}
% Bibliography style - if using a .bib file
\bibliographystyle{IEEEtran}
\bibliography{reference1} % without .bib extension

% that's all folks
\end{document}